# High resolution spectra of the [6297-6303] and [6361-6367] Angström domains (including forbidden OI lines) of the Sun and brightest stars

*Jean-Marie Malherbe (emeritus astronomer)*

Observatoire de Paris, PSL Research University, LIRA, France

Email: Jean-Marie.Malherbe@obspm.fr; ORCID: https://orcid.org/0000-0002-4180-3729

4 March 2025

**ABSTRACT**

We present a dataset of high resolution spectra of the Sun and ten bright stars of the domains [6297-6303] and [6361-6367] Angtröm. Solar spectra were obtained in the quiet Sun at various distances from disk centre with the ground based Meudon Solar Tower and Themis telescope (12 mÅ resolution) and with the Solar Optical Telescope (SOT) onboard the Hinode satellite (21 mÅ resolution). Spectra of 10 bright stars (magnitude < 2) were also got with Themis at 12 mÅ resolution. These spectral domains contain the faint and forbidden OI lines (6300.31 Å and 6363.79 Å) that are useful for the research of Oxygen abundance. The spectra shown here are freely available in FITS format to the research community.

**INTRODUCTION**

The spectroscopic determination of the abundance of oxygen in the solar photosphere is not an easy task: few atomic lines are available in the solar spectrum, and most of them are mixed with lines of other elements (blends such as NiI at 6300.4 Å). A great deal of effort has been devoted to the spectroscopic determination of the oxygen abundance, without any convergence on a definitive value, as discussed by Caffau *et al* (2008, 2013, 2015, 2017) and Steffen *et al* (2015). The datasets presented here were used by Caffau *et al* (2013, 2015) ; we offer them now on-line to be used freely by the research community.

**OBSERVATIONS**

For Hinode it was non standard observations, because usually the 6300 window is centred on FeI lines at 6301.5 Å and 6302.5 Å which are Zeeman sensitive and very useful to study the solar magnetism and activity. For our observation, the window was slightly decentred towards the blue in order to record the forbidden OI line at 6300.3 Å (usually out of band) ; the spectral resolution is R = 200000. For Themis, the observation of bright stars is also not standard, since it is a solar telescope. However, its 90 cm diameter allows to produce detailed spectra of the brightest stars (magnitude < 2) with the solar spectrograph (R = 300000). The observing procedure with Hinode and Themis is described in details by Caffau *et al* (2015). At Meudon, we used the large spectrograph of the Solar Tower which has a spectral resolution R = 400000.

The table below summarizes the nature of observations and in particular gives the number of observations for each spectral domain (both the 6300 and 6363 columns) ; in the case of solar observations, various $\mu = \cos(\theta)$ locations were chosen along the equator ($\mu$ column).

| Star | Apparent Mag | Spectral type | R/Rs | T (K) | 6300 | 6363 | μ values | sampling (mÅ) |
|---|---|---|---|---|---|---|---|---|
| Antares | 1,1 | M1.5 | 800 | 3600 | 1 | | | 12 |
| Betelgeuse | 0,0 | M1-2 | 1000 | 3600 | 1 | 1 | | 12 |
| Aldebaran | 0,9 | K5 | 45 | 3900 | 1 | 1 | | 12 |
| Arcturus | 0,0 | K1.5 | 25 | 4300 | 4 | 2 | | 12 |
| Hamal | 2,0 | K2 | 15 | 4500 | 1 | | | 12 |
| Diphda | 2,0 | K0 | 17 | 4800 | 1 | | | 12 |
| Pollux | 1,1 | K0 | 9 | 4900 | 1 | 1 | | 12 |
| Capella AB | 0,8 | G0/G5 | 10 | 5300 | 2 | 2 | | 12 |
| Procyon | 0,4 | F5 | 2 | 6550 | 4 | | | 12 |
| AlphaPersei | 1,8 | F5 | 68 | 6350 | 1 | | | 12 |
| SUN (Themis) | -26 | G2 | 1 | 5800 | 2 | 1 | 8 | 12 |
| SUN (Hinode) | -26 | G2 | 1 | 5800 | 2 | | 5 | 21 |
| SUN (Meudon) | -26 | G2 | 1 | 5800 | 1 | | 10 | 12 |

Solar spectral lines

Range [6297-6303]

```
Wavelength  Eq. Width  Red.     Element
(A)         (mA)       width
6297.262      7.        1.       ATM H2O
6297.799S    65.       11.S      FE 1
6298.084R     5.        1.       TI 1"
6298.457M    22.        4.       ATM O2
6299.228S    30.        5.       ATM O2
6299.228S    30.        5.       ATM H2O
6300.311      5.        1.       ((O 1))
6300.678      6.        1.       SC 2
6301.508S   127.       19.S"     FE 1
6301.845R     2.        0.       FE 1P
6302.000M    23.        4.       ATM O2
6302.499S    83.       13.W      FE 1
6302.764S    21.        3.       ATM O2

6361.252 *    5.        1.       ATM H2O
6361.940     89.       14.       CA 1
6362.350M    23.        4.W      ZN 1
6362.876M    30.        5.SS     //CR 1
6362.876M    30.        5.SS     FE 1
6363.790M     3.        1.       ((O 1))
6364.369M    25.        4.W"     FE 1
6364.706     12.        2.W      FE 1
6366.356      4.        1.SS     TI 1
6366.491M    26.        4.W      NI 1
```

Range [6360-6366]

**AVAILABLE DATASETS**

All data (wavelengths and intensities) are included in FITS format in the joint 6300-6363.zip file, which is freely available with this document, and also archived at : DOI 10.6084/m9.figshare.28497671 or https://figshare.com/s/da9705ab043661700cd4

Quiet Sun observations are archived in directories "sunMeudon6300", "sunHinode6300", "sunThemis6300" and "sunThemis6363".

Three sets were produced by Themis, two for lines around 6300 Å and one for lines around 6363 Å. For each value of μ, 2D spectra images (2D files "img6301mu*.fits" or "img6364mu*.fits" with λ in the first direction and the abscissa along the slit in the second direction), together with space averaged line profiles (1D files "6301mu*.fits" or "6364mu*.fits") are available. The 2D spectra images result from time averaging sequences of 350 to 400 exposures at constant μ. The spectral resolution is 11.5 mÅ (6363 Å) or 12 mÅ (6300 Å). μ values are 1.0, 0.96, 0.88, 0.72, 0.66, 0.48, 0.36, 0.25.

Two sets were produced by Hinode for lines around 6300 Å (three lines visible, OI, ScII and FeI with 21.5 mÅ spectral resolution). For each value of µ, 2D spectra images (2D files "OI-mu*.fits" with λ in the first direction and the abscissa along the slit in the second direction) and space averaged line profiles (1D files "OImean-mu*.fits" and "OIavg-mu*.fits") are available. "OIavg-mu*.fits" is the average of two "OImean-mu*.fits" files, at the same µ, but alternatively located in the Eastern or western hemisphere, along the solar equator. µ values are 0.375, 0.692, 0.856, 0.950, 0.994 (East side) and 0.387, 0.697, 0.859, 0.952, 0.994 (West side).

One set was produced by Meudon for lines around 6300 Å (three lines visible, OI, ScII and FeI with 11.6 mÅ spectral resolution). For each value of µ, 2D spectra images (2D files "spimg-mu*.fits" with λ in the first direction and the abscissa along the slit in the second direction) and space averaged line profiles (1D files "tour-mu*.fits") are available. Some parasitic interference fringes are present in the spectra images, but have been partly removed in average profiles. µ values are 1.0, 0.96, 0.88, 0.72, 0.66, 0.48, 0.35, 0.25, 0.20, 0.15.

Concerning bright stars, many spectra around 6300 Å and 6363 Å were produced by Themis with the solar spectrograph (with the resolution of 12 mÅ). However, there is more noise because Themis is a small telescope (90 cm aperture) ; the files are the result of time averaging sequences of 30 to 60 exposures. The directories are of the form "star6300" or "star6300-6363", when the two spectral domains were recorded, and the spectra are named "star6300.fits" or "star6363.fits". For some stars, there is a video of the slit jaw when available.

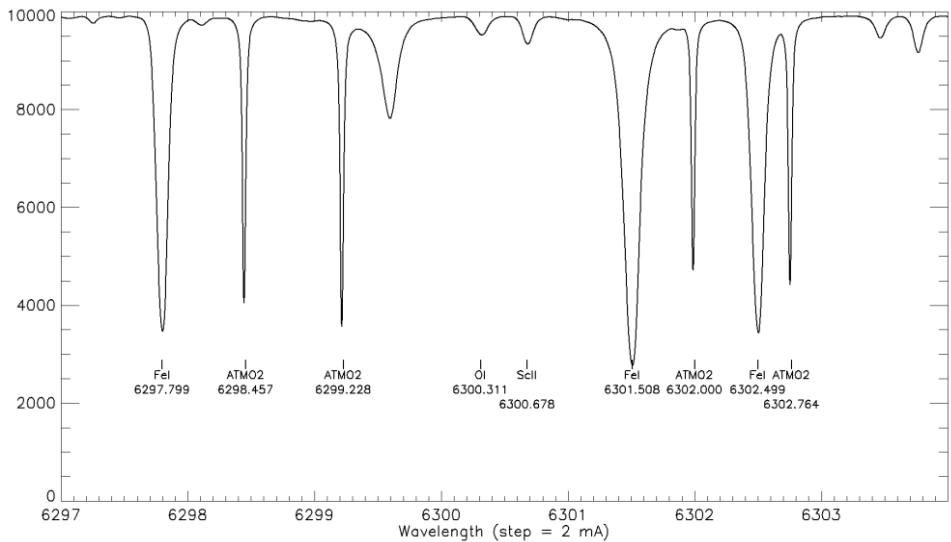

*Solar atlas, from Delbouille et al (1972), at disk centre, spectral window at 6300 Å, wavelength resolution 0.002 Å*

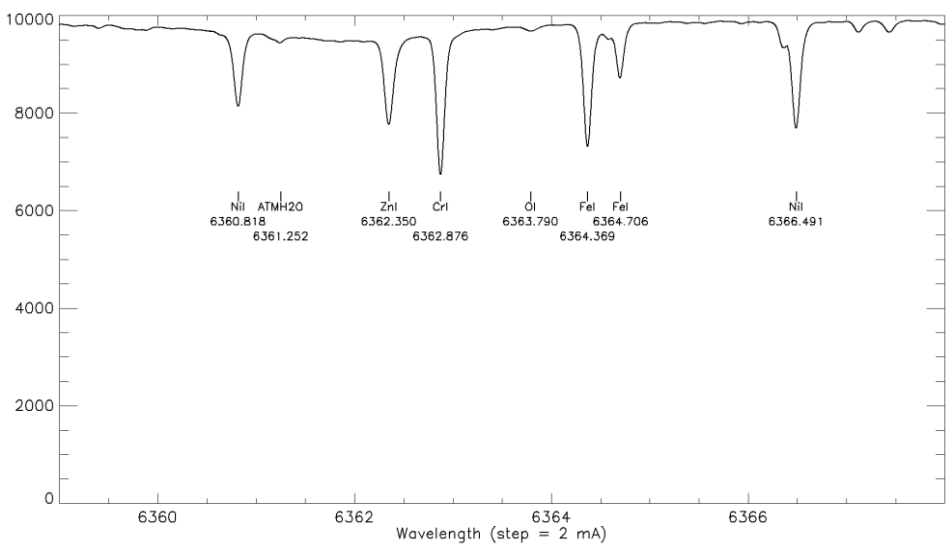

*Solar atlas, from Delbouille et al (1972), at disk centre, spectral window at 6363 Å, wavelength resolution 0.002 Å*

**SOLAR DATASETS : some examples of recent observations (figures derived from the datasets).**

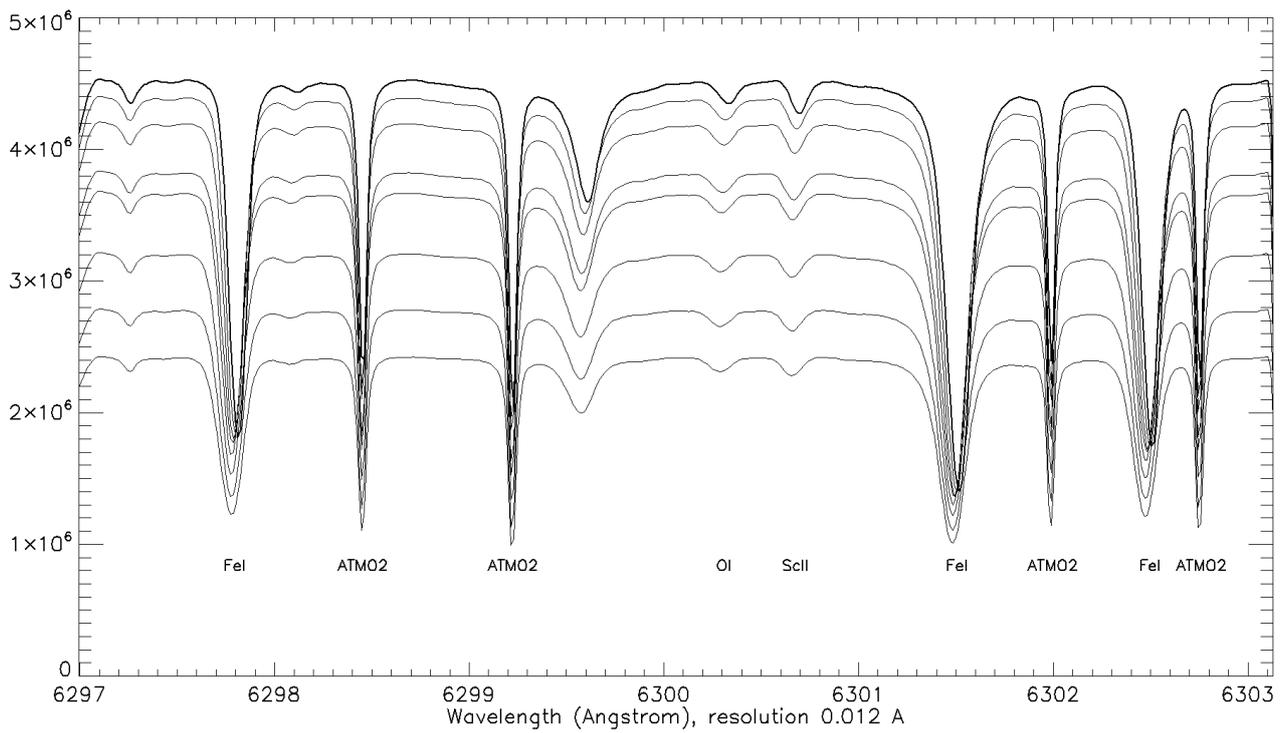

*Sun, Themis, spectral window 6300 Å, wavelength resolution 0.012 Å, 25 April 2011, 18:01-18:28 UT (set 1), and 26 April 2011, 15:42-16:01 UT (set 2), both for µ = 1.0, 0.96, 0.88, 0.72, 0.66, 0.48, 0.36, 0.25*

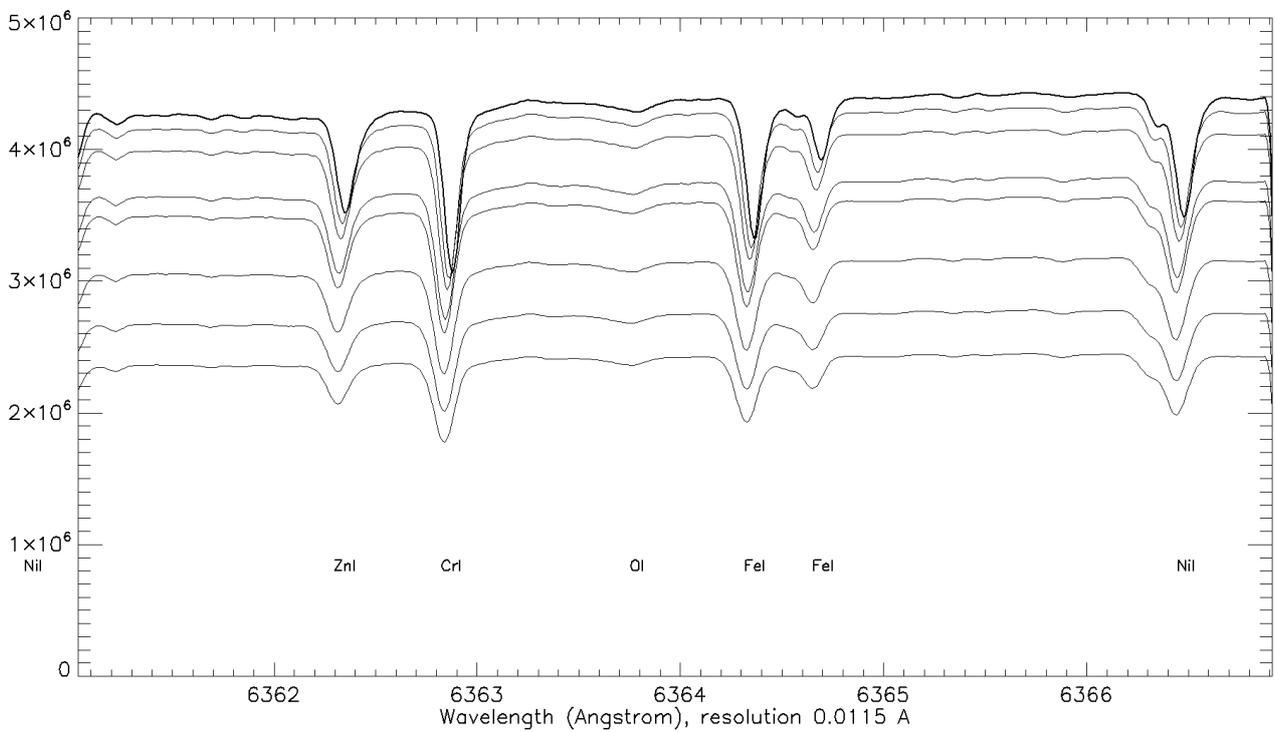

*Sun, Themis, spectral window 6363 Å, wavelength resolution 0.0115 Å, 26 April 2011, 16:15-16:49 UT, for µ = 1.0, 0.96, 0.88, 0.72, 0.66, 0.48, 0.36, 0.25*

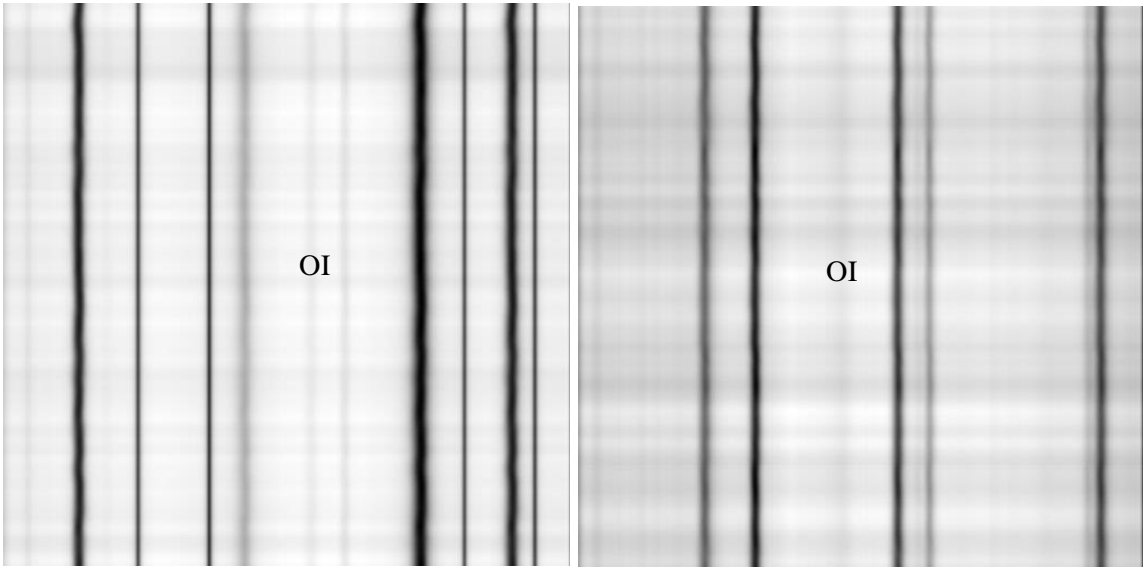

*Sun, Themis, 2D (λ, x) spectral windows at 6300 Å (left) and 6363 Å (right), wavelength resolution 0.0115 Å, 25/26 April 2011, for µ = 1.0 (disk centre). Other values of µ (0.96, 0.88, 0.72, 0.66, 0.48, 0.36, 0.25) are available in the datasets.*

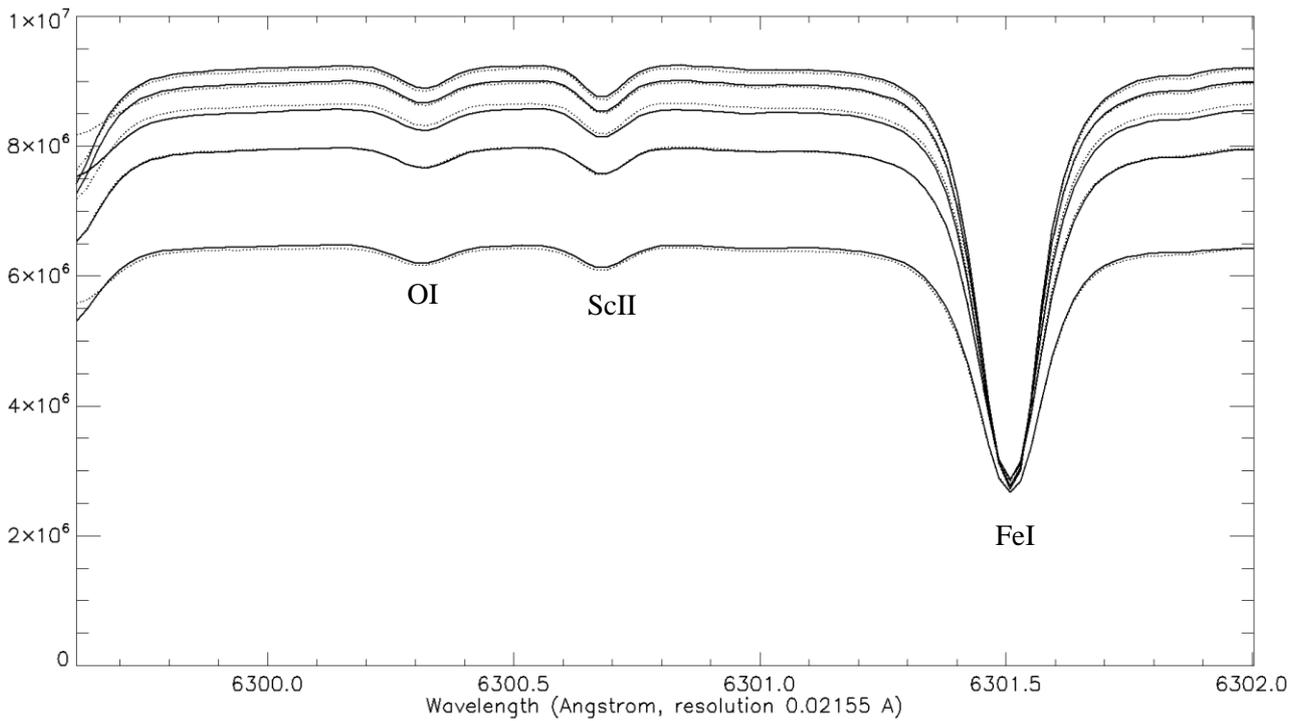

*Sun, Hinode, spectral window 6300 Å, wavelength resolution 0.02155 Å, 10 November 2010, 19:26-23:07 UT, for µ = 0.375, 0.692, 0.856, 0.950, 0.994 (along the equator, East side, solid line) and for µ = 0.387, 0.697, 0.859, 0.952, 0.994 (along the equator, West side, dotted lines). An average of two positions at same µ is also provided. Telluric lines are of course absent. Original data were obtained in polarimetric mode I+V, I-V suitable for circular analysis of the Zeeman effect. We used only intensities.*

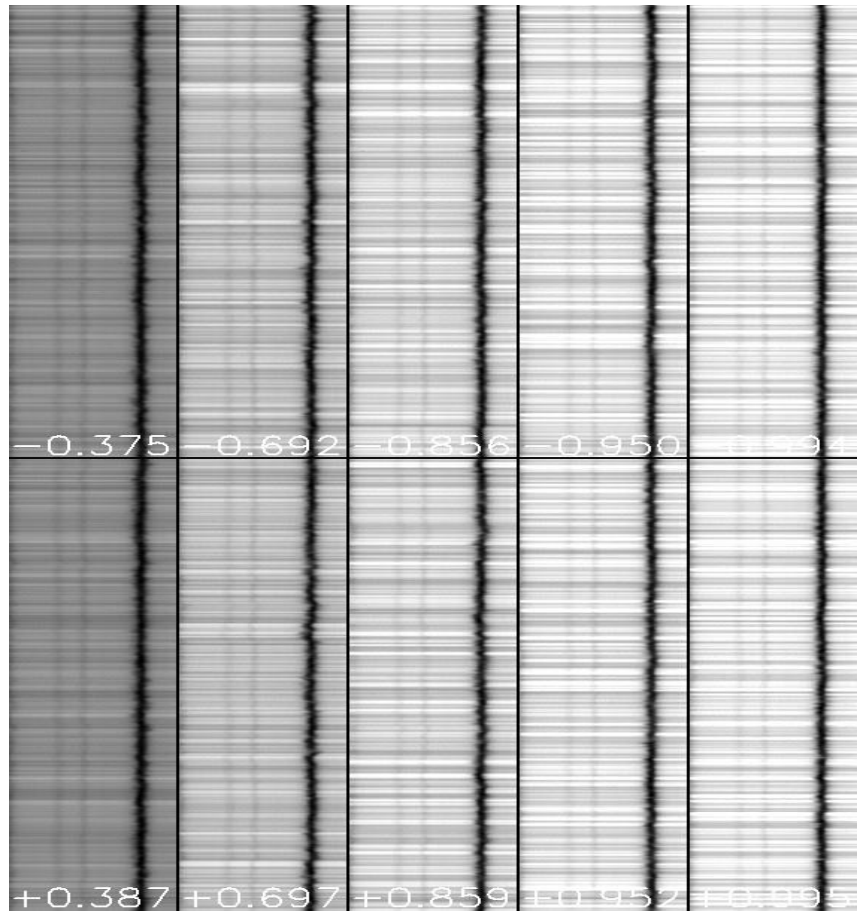

*Sun, Hinode, 2D (λ, x) spectral window 6300 Å, wavelength resolution 0.02155 Å, 10 November 2010, 19:26-23:07 UT, for μ = 0.375, 0.692, 0.856, 0.950, 0.994 (along the equator, East side, minus sign) and for μ = 0.387, 0.697, 0.859, 0.952, 0.994 (along the equator, West side)*

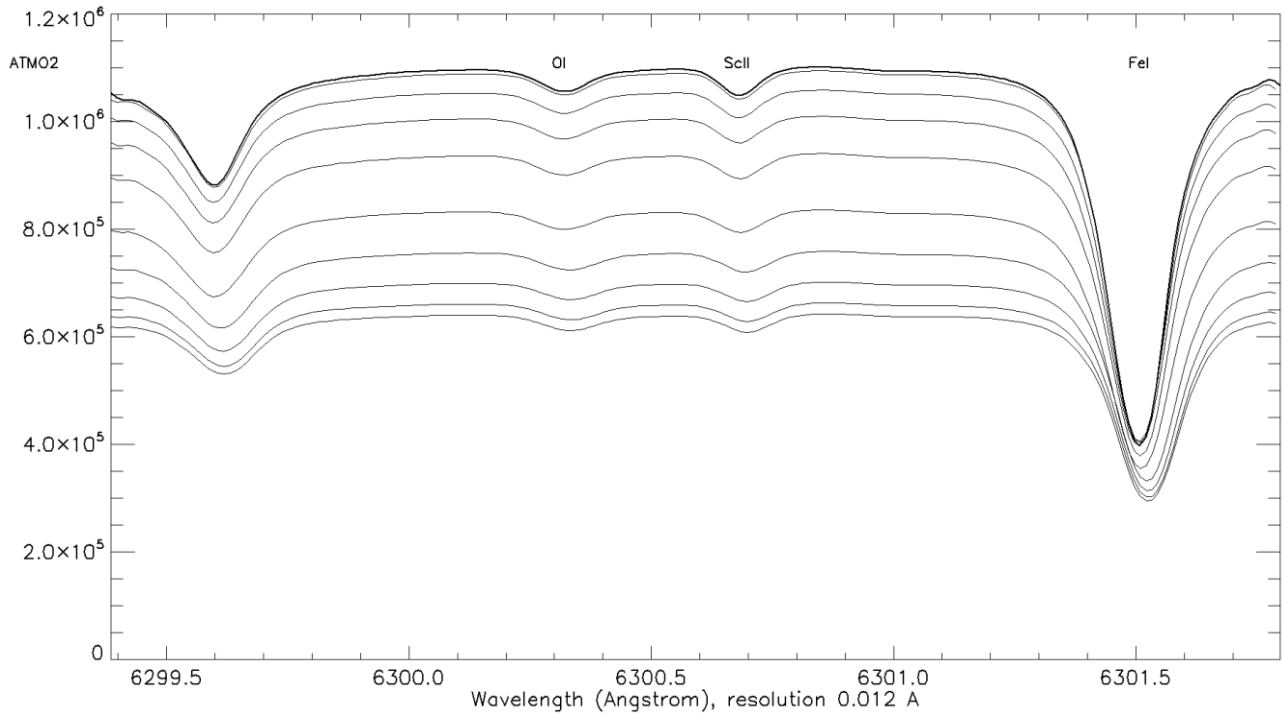

*Sun, Meudon, spectral window 6300 Å, wavelength resolution 0.0116 Å, 25 May 2011, 10:36-11:29 UT, for μ = 1.0, 0.96, 0.88, 0.72, 0.66, 0.48, 0.35, 0.25, 0.20, 0.15. Profiles are not corrected from the rotational Doppler effect.*

# BRIGHT STARS DATASETS : some examples of recent observations (figures derived from the datasets)

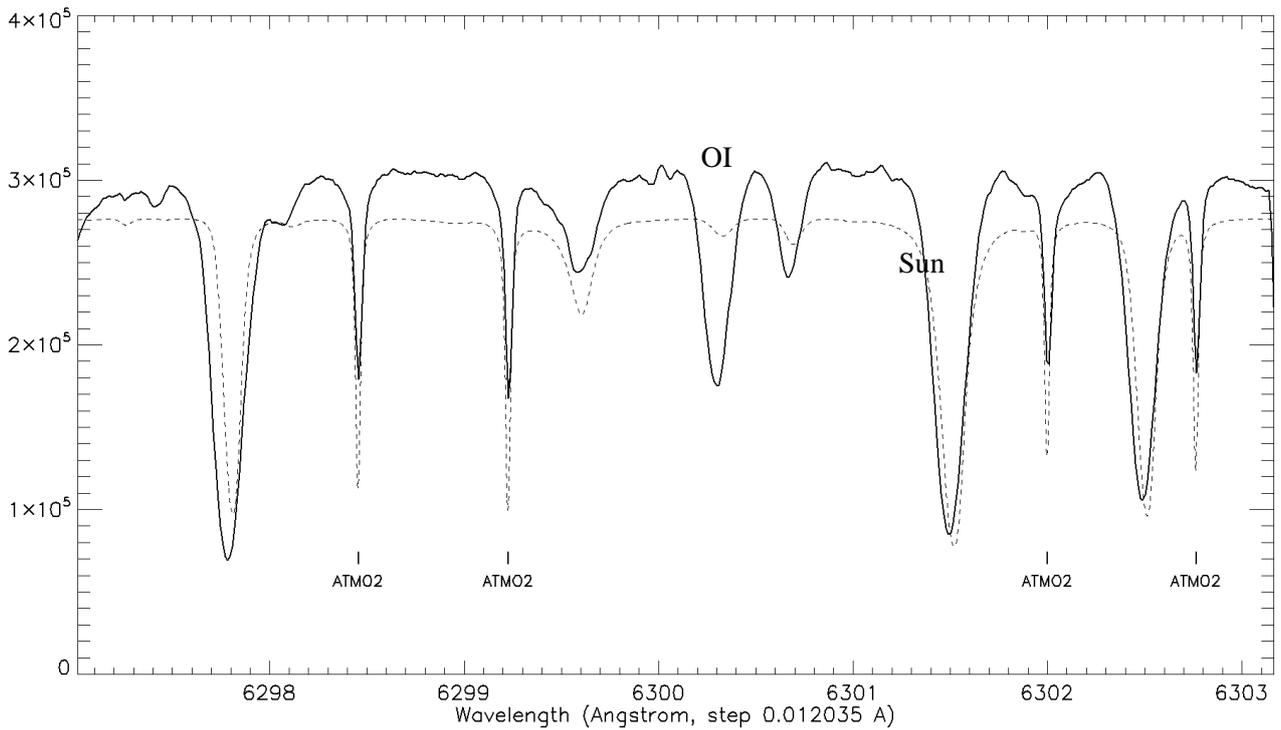

*Arcturus, spectral window 6300 Å, wavelength resolution 0.012 Å, 26 April 2011, 00:52 and 23:44 UT (SET A, two observations)*

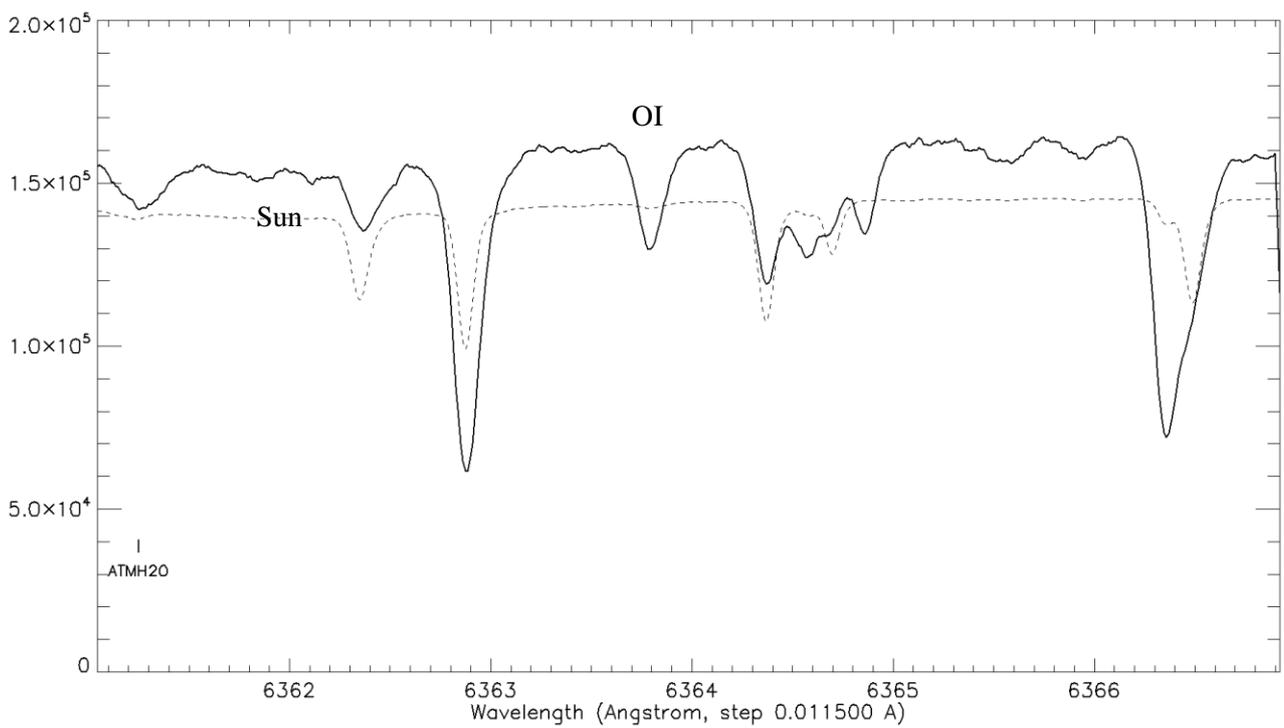

*Arcturus, spectral window 6363 Å, wavelength resolution 0.0115 Å, 27 April 2011, 00:52 UT (SET A)*

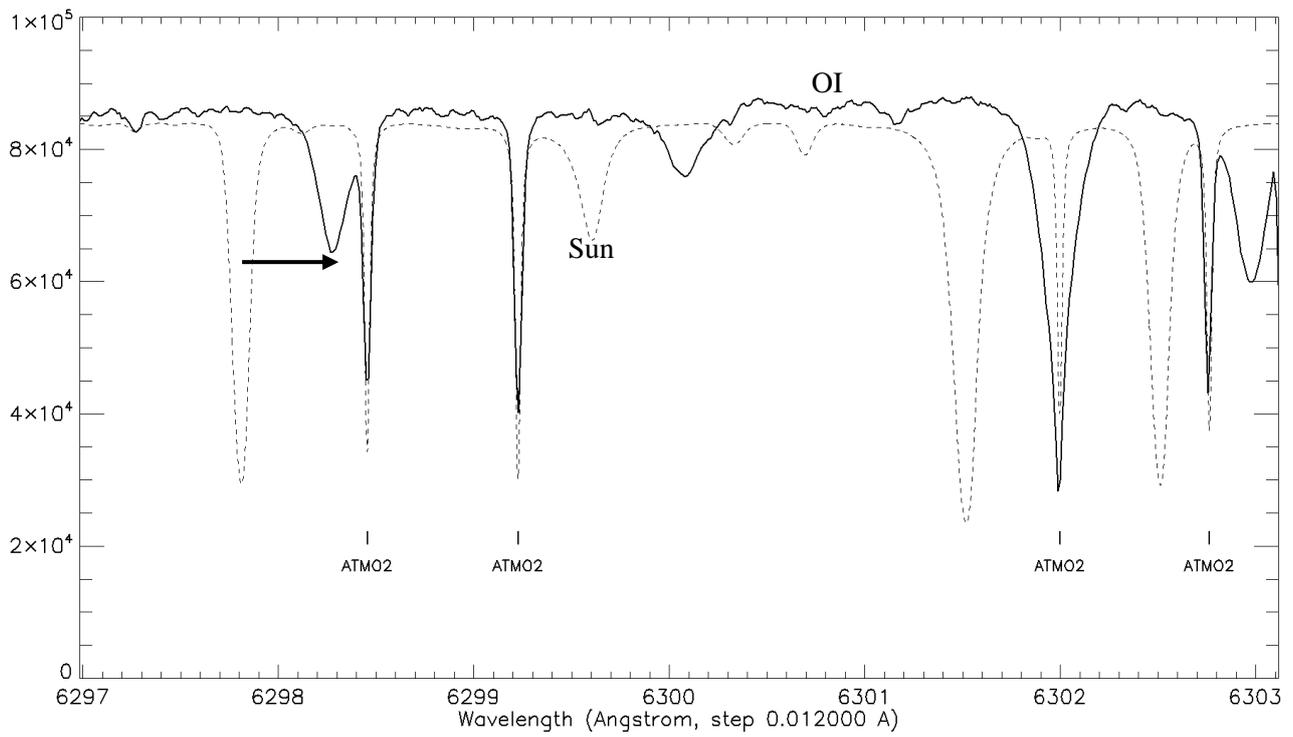

*Procyon, spectral window 6300 Å, wavelength resolution 0.012 Å, 25 April 2011, 21:44 UT (set 1), 26 April 2011, 21:21 UT (set 2), 25 April 2012, 23:22 UT (set 3) and 25 April 2012, 23:22-01:08 UT for set 4.*

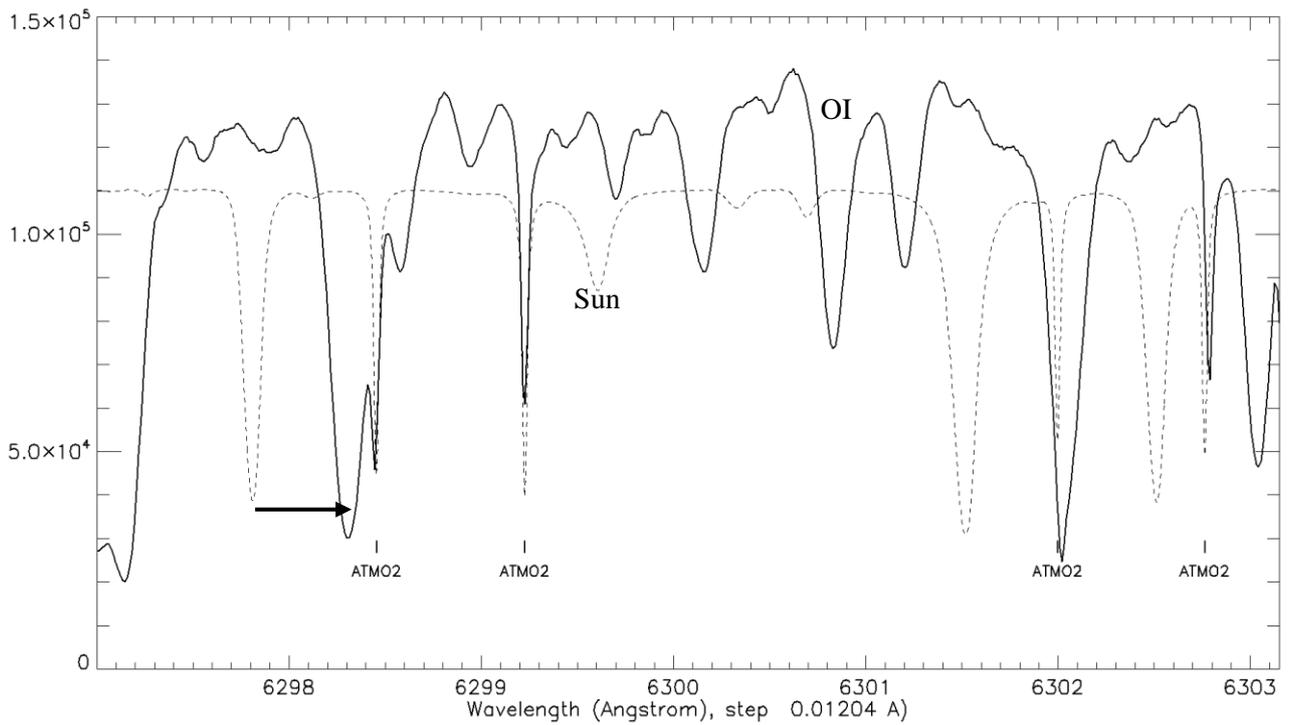

*Aldebaran, spectral window 6300 Å, wavelength resolution 0.012 Å, 7 September 2011, 03:31 UT*

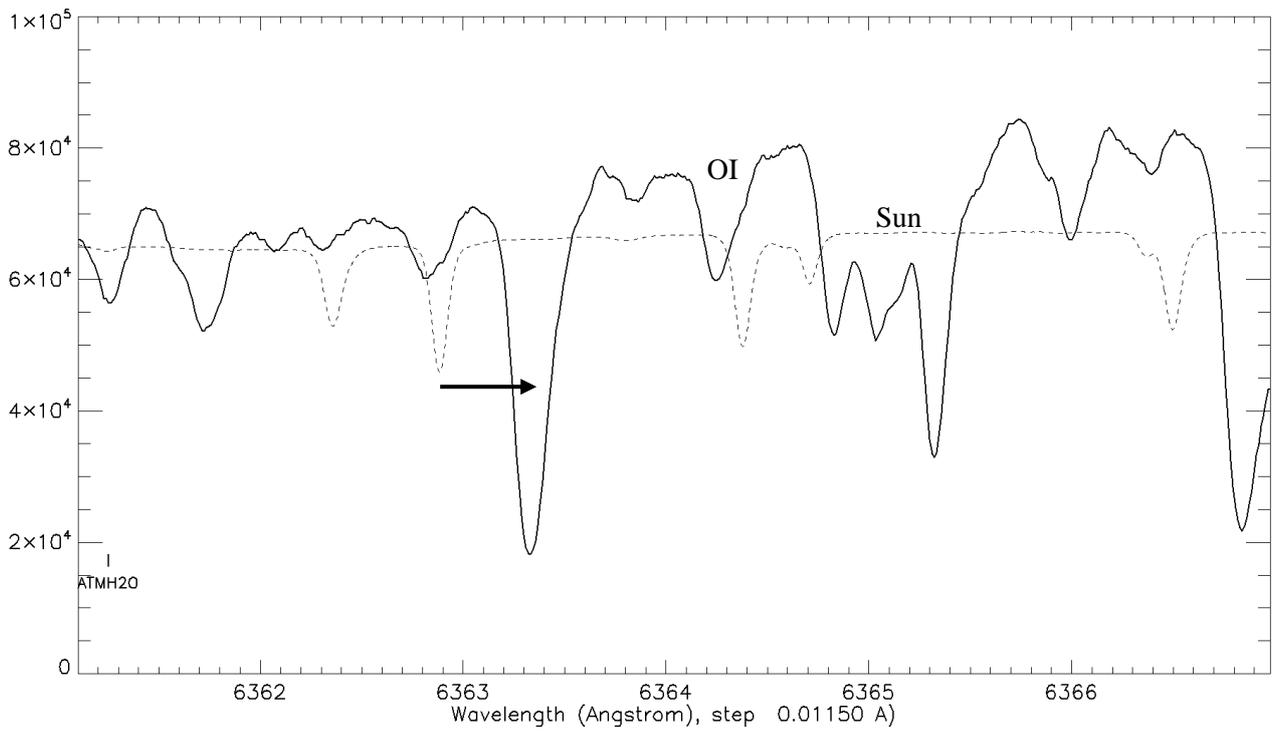

*Aldebaran, spectral window 6363 Å, wavelength resolution 0.0115 Å, 7 September 2011, 04:36 UT*

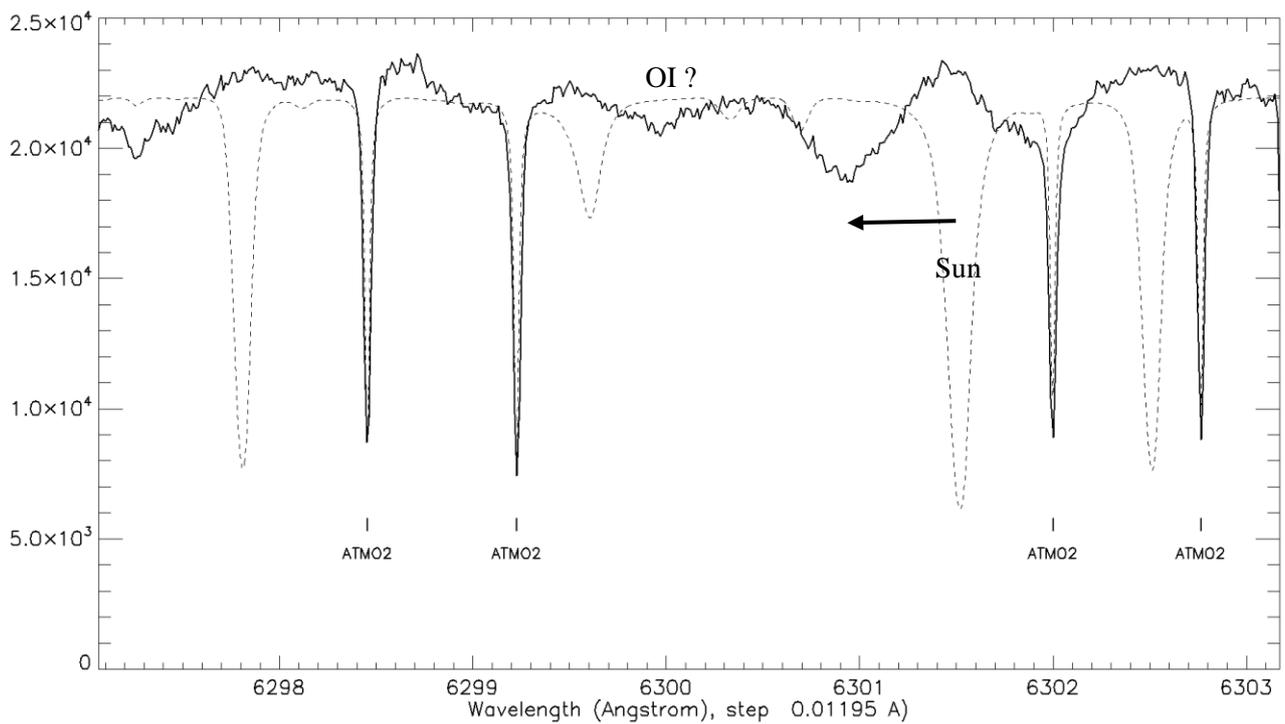

*AlphaPersei, spectral window 6300 Å, wavelength resolution 0.01195 Å, 6 September 2011, 23:51 UT, FeI lines are very faint.*

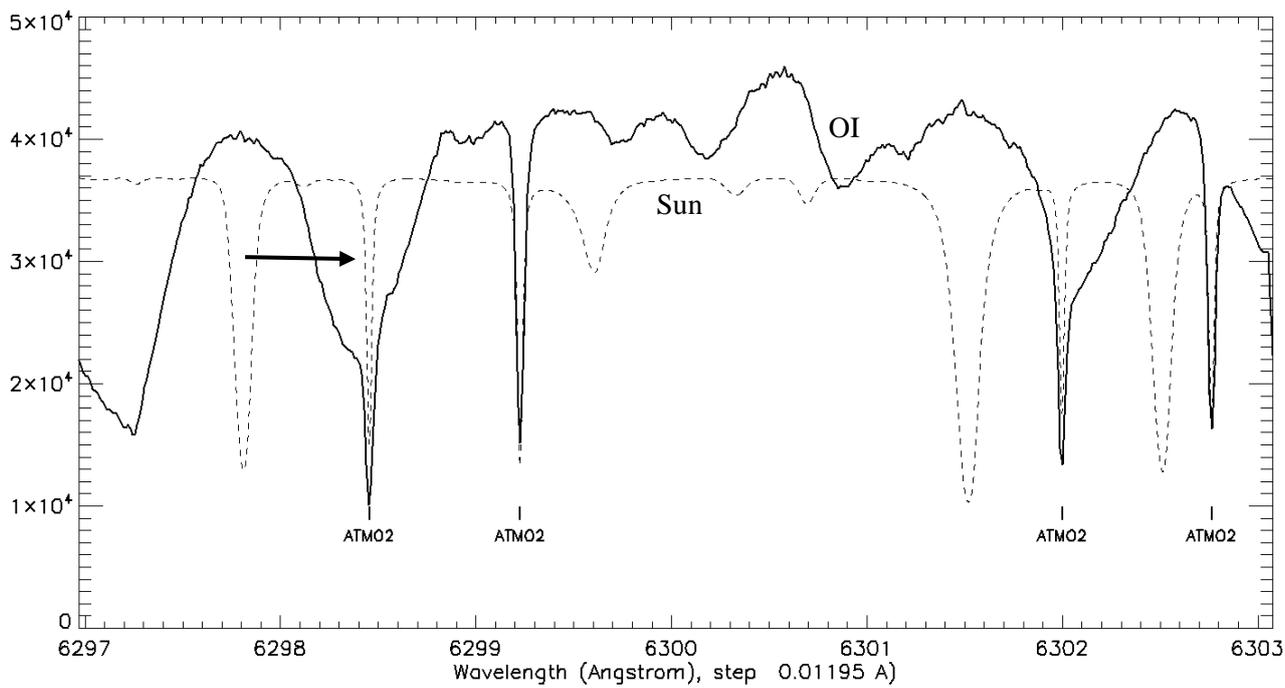

*Antares, spectral window 6300 Å, wavelength resolution 0.01195 Å, 7 September 2011, 20:14 UT, FeI lines are very large.*

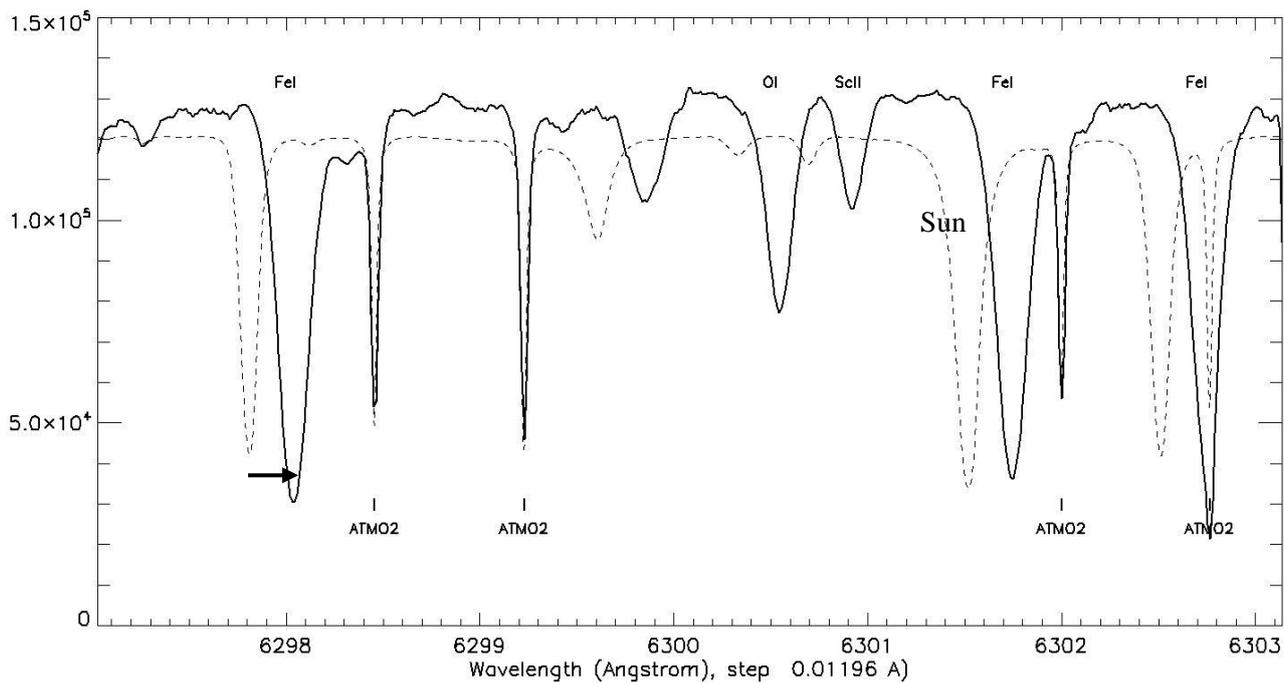

*Arcturus, spectral window 6300 Å, wavelength resolution 0.012 Å, 6 September 2011, 20:18 and 21:23 UT (SET B, two observations)*

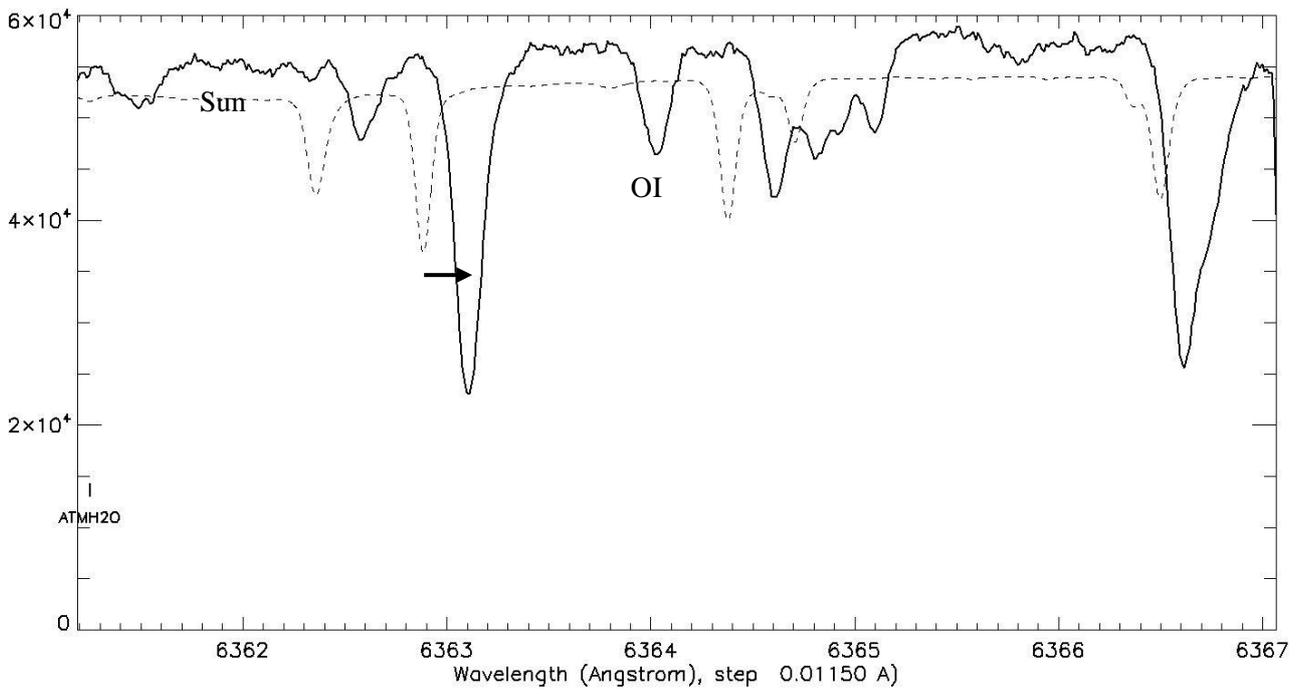

*Arcturus, spectral window 6363 Å, wavelength resolution 0.0115 Å, 6 September 2011, 21:22 UT (SET B)*

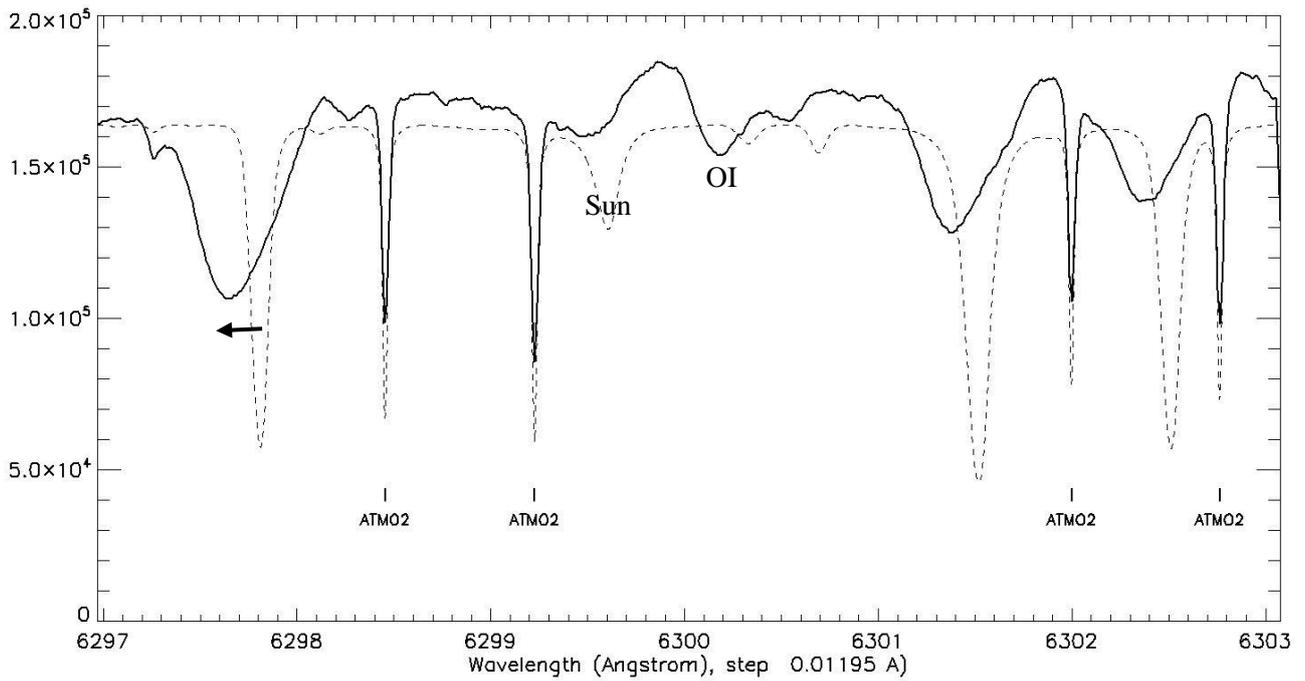

*Betelgeuse, spectral window 6300 Å, wavelength resolution 0.012 Å, 7 September 2011, 05:43 UT*

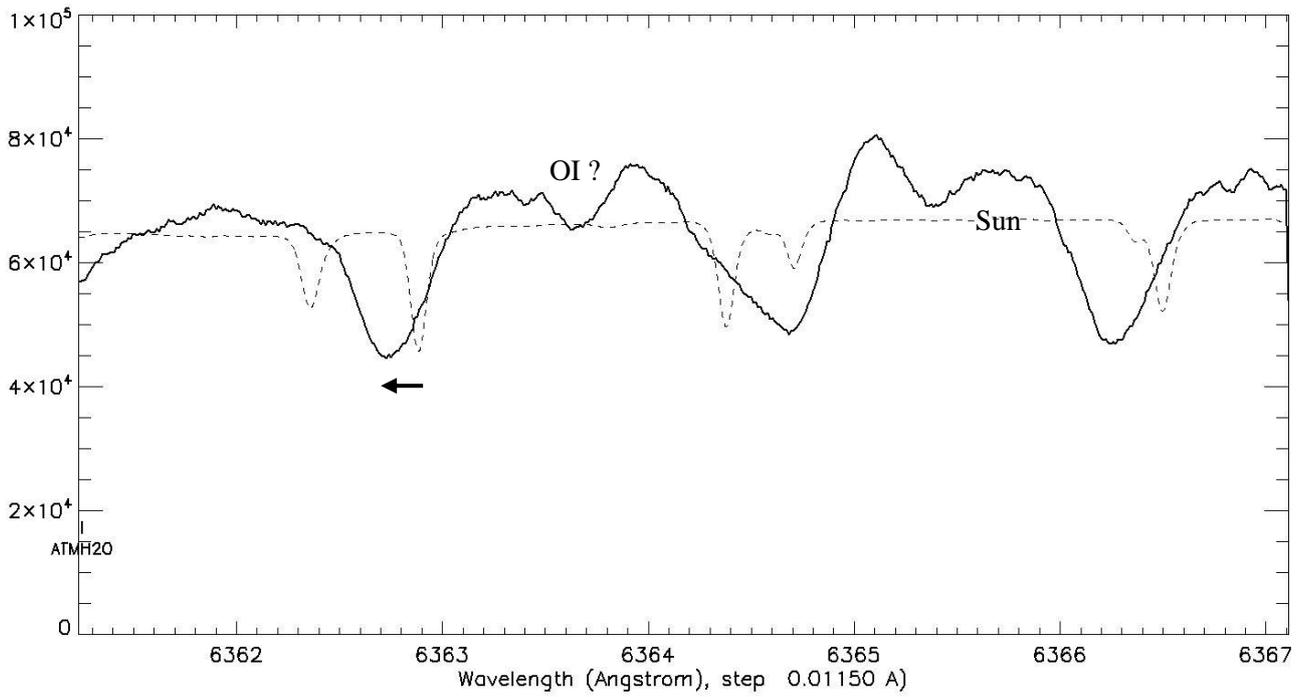

*Betelgeuse, spectral window 6363 Å, wavelength resolution 0.0115 Å, 8 September 2011, 03:48 UT*

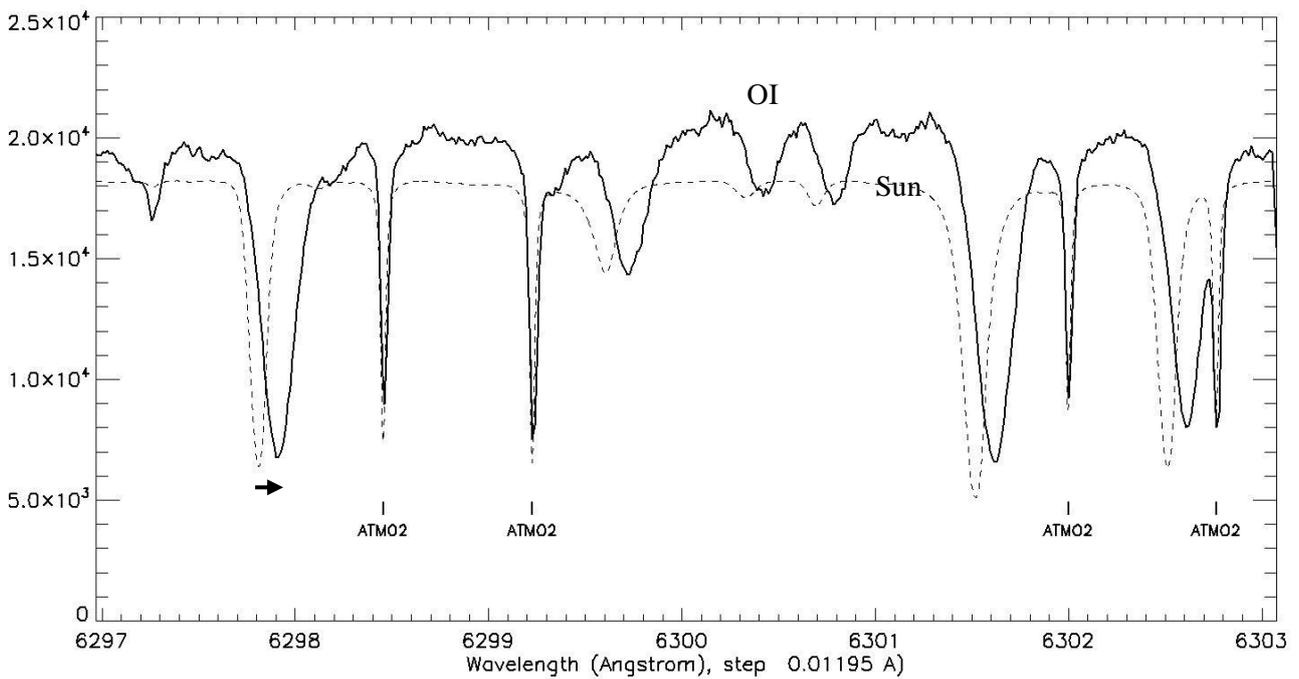

*Diphda, spectral window 6300 Å, wavelength resolution 0.012 Å, 7 September 2011, 23:47 UT*

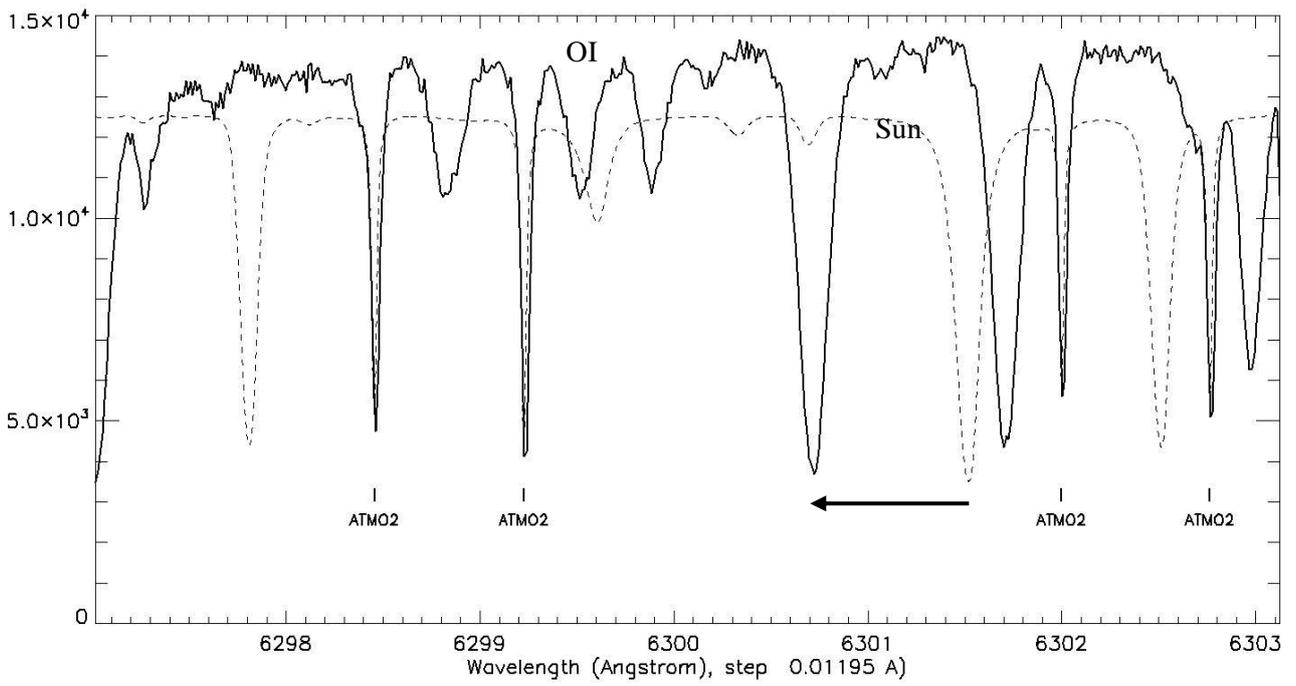

*Hamal, spectral window 6300 Å, wavelength resolution 0.012 Å, 7 September 2011, 22:40 UT*

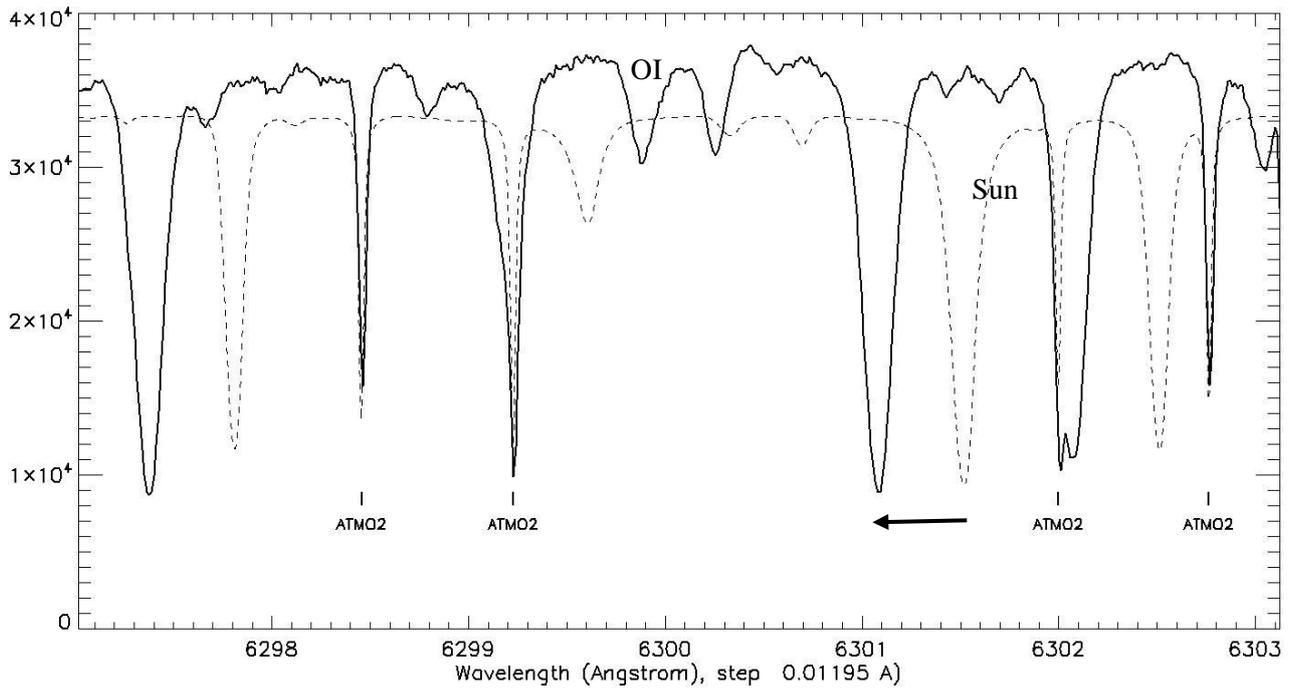

*Pollux, spectral window 6300 Å, wavelength resolution 0.012 Å, 8 September 2011, 04:58 UT*

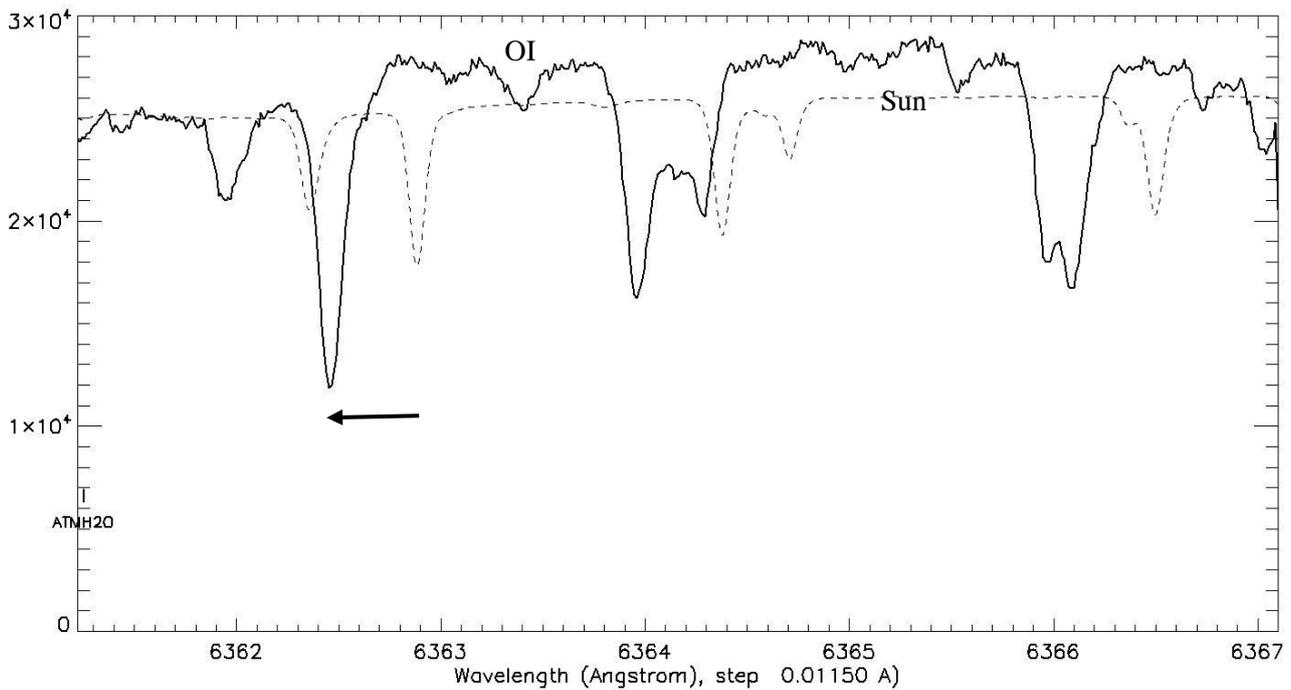

*Pollux, spectral window 6363 Å, wavelength resolution 0.0115 Å, 8 September 2011, 06:01 UT*

**REFERENCES**

E. Caffau1, H.-G. Ludwig, M. Steffen, T. R. Ayres, P. Bonifacio, R. Cayrel, B. Freytag, and B. Plez "The photospheric solar oxygen project - I. Abundance analysis of atomic lines and influence of atmospheric models", A&A, 488, 1031–1046 (2008)

E. Caffau1, H.-G. Ludwig, J.-M. Malherbe, P. Bonifacio, M. Steffen, and L. Monaco, "The photospheric solar oxygen project - II. Non-concordance of the oxygen abundance derived from two forbidden lines", A&A, 554, A126 (2013)

E. Caffau, H.-G. Ludwig, M. Steffen, W. Livingston, P. Bonifacio, J.-M. Malherbe, H.-P. Doerr, and W. Schmidt, "The photospheric solar oxygen project - III. Investigation of the centre-to-limb variation of the 630nm [O I]-Ni I blend", A&A, 579, A88 (2015)

M. Steffen, D. Prakapavicius, E. Caffau, H.-G. Ludwig, P. Bonifacio, R. Cayrel, A. Kucinskas, and W. C. Livingston, "The photospheric solar oxygen project - IV. 3D-NLTE investigation of the 777nm triplet lines",_A&A, 583, A57 (2015)

E. Caffau, J.-M. Malherbe, M. Steffen, H.-G. Ludwig, and A. Mott, "Investigation of the solar centre-to-limb variation of oxygen and lithium spectral features", Mem. S.A.It. Vol. 88, 45 (2017)